\documentstyle[epsf,bibnorm]{lamuphys}
\makeatletter
\let\chapter\hid@chapter
\makeatother
\def\be{\begin{equation}}
\def\ee{\end{equation}}
\def\bea{\begin{eqnarray}}
\def\eea{\end{eqnarray}}

\begin{document}
\pagenumbering{arabic}

\title{Nucleon Resonances in Kaon Photoproduction}

\author{C. \,BENNHOLD\inst{1}, T. MART\inst{1,2}, A. WALUYO\inst{1}, 
H. HABERZETTL\inst{1}, G. PENNER\inst{3}, T. FEUSTER\inst{3}, U. MOSEL\inst{3}}

\institute{Center for Nuclear Studies, Department of Physics,
         The George Washington University, Washington, 
         D.C. 20052, U.S.A.
\and
Jurusan Fisika, FMIPA, 
         Universitas Indonesia, Depok 16424, Indonesia
\and
Institut f\"ur Theoretische Physik, Universit\"at Giessen,
         D-35392 Giessen, Germany}

\maketitle
\begin{abstract}
        Nucleon resonances are investigated through 
        the electromagnetic production of K-mesons.
        We study the kaon photoproduction process 
        at tree-level and compare to a recently developed unitary
        K-matrix approach.  Employing hadronic form factors
        along with the proper gauge prescription yields 
        suppression of the Born terms and leads a resonance dominated
        process for both $K \Lambda$ and $K \Sigma$ photoproduction.
        Using new SAPHIR data we find the $K^+ \Lambda$ photoproduction
        to be dominated by the $S_{11}$(1650) at threshold, with 
        additional contributions from the $P_{11}$(1710) and
        $P_{13}$(1720) states. The $K \Sigma$ channel couples
        to a cluster of $\Delta$ resonances around $W$ = 1900 MeV.
        We briefly discuss some tantalizing evidence for a missing $D_{13}$ 
        resonance around 1900 MeV with a strong branching ratio
        into the $K \Lambda$ channel.
\end{abstract}
  
\section{Introduction}
 
Since its discovery five decades ago, the quark flavor strangeness
has played a special role in nuclear physics and particle physics.
More recently, the strange quark has found itself between two theoretical
domains: on the one hand is the
realm of chiral symmetry with the almost massless up 
and down quarks, while on the other side the physics can be 
described in terms of the heavy quark effective theory of the charm and
bottom quarks.  While the strange quark mass is too large to
ensure convergence in SU(3) Chiral Perturbation Theory, it turns out
to be too small to be safely included in heavy quark descriptions.

Among the successes of heavy quark physics is the straightforward
description of excitation spectra of mesons and baryons that contain
charm and bottom quarks.  The excitation spectrum of nucleon and
hyperon resonances, on the other hand, is still not well understood, despite 40 years
of efforts in meson-baryon scattering and electromagnetic production processes.
For this purpose, a number of laboratories like MAMI, ELSA, BATES, GRAAL
and TJNAF have begun to address the issue of $N^*$ physics,
delivering new experimental data with unprecedented accuracy.

On the theoretical side, progress is being made in the
understanding of $N^*$ and $Y^*$
properties from first principles calculations, such as 
lattice QCD which provides
numerical solutions of QCD on a discrete space-time lattice.
Due to the improved actions in the last few years, coupled with advances in 
algorithms and computing power, the field is quickly moving 
towards providing new results in hadron phenomenology. 

In order to provide a link between the new and improved data on one side 
and the results from lattice QCD on the other side,
dynamical descriptions using hadronic degrees of freedom
 are required that can analyze the data in the
 various asymptotic reaction channels (like $\gamma N, \pi N, \pi \pi N,
\eta N, K \Lambda, K \Sigma$ and others).
In order to preserve unitarity and analyticity these analyses
need to be performed in a unitary coupled-channels framework which allows
separating background from resonance contributions.
A number of different approaches have been developed over the last two 
decades \cite{ben:cutkosky79,ben:hoehler79,ben:manley92,ben:vpi95,ben:feuster98}, 
here we focus on the description of resonances within the effective Lagrangian framework.

\section{The $K$-Matrix Born Approximation}

Within the one-photon approximation, the full amplitude for any photoproduction process can be 
written in terms of a Bethe-Salpeter equation
\begin{eqnarray}
M ~ = ~V ~+ ~ V~G~T,
\label{eq:tmatrix}
\end{eqnarray} 
where $V$ represents the driving term for the particular photoproduction
process, $G$ is the meson-baryon two-particle propagator, and $T$ is the
hadronic meson-baryon final state interaction.  In principle, one would
have to solve this equation as a four-dimensional integral equation. However
in practice, due to the singularity structure involved,
 this has only recently been pursued for the case of pion scattering
with a small number of diagrams in the driving term. Generally,
a three-dimensional reduction is chosen that amounts to
making an assumption of the intermediate two-particle propagator
which then makes the calculations more tractable. Writing the full 
Bethe-Salpeter equation in the form 
\begin{eqnarray}
K &=& V + V~{\rm Re}(G_{\rm BS})K\\
T &=& K + iK~{\rm Im}(G_{\rm BS})T ~.
\end{eqnarray}
where $G_{BS}$ is the full propagator, then any truncation of the first 
equation will still provide a unitary, albeit approximate, solution, as long as 
$i Im(G_{BS})$ correctly describes the discontinuity across the scattering 
cut. Taking the special choice
\begin{eqnarray}
i G_{\rm BS} &=& -2i(2\pi)^2m_N\delta(k_N^2-m_N^2)\delta(k_m^2-m_m^2)\nonumber\\
           && \times ~\theta (k_N^0)\theta (k_m^0)(k\!\!\!/_N +m_N) ~,
\end{eqnarray}
and $K=V$ leads to the simple K-matrix Born approximation:
\begin{eqnarray}
T &=& \frac{V}{1-iV} ~.
\end{eqnarray}

Thus, both intermediate particles are being placed on-shell.  This procedure
still allows for the resonance widths to be generated dynamically, while
the real part of the self-energy is absorbed in an effective resonance mass
that is determined by the fit.
The most recent coupled-channels approach that employs the $K$-matrix
approximation within an effective Lagrangian framework has been performed
by Feuster and Mosel \cite{ben:feuster98}. 
They extract nucleon resonance parameters by simultaneously analyzing
all available data for reactions involving the initial and final states
$\gamma N, \pi N, \pi \pi N, \eta N$ and $K \Lambda$ up to 
$W = 1.9$ GeV.

\section{Kaon Photoproduction in the Coupled-Channels Approach}

While dynamical models involving various
approximations for the Bethe-Salpeter equation
are becoming increasingly successful in the description 
of pion photoproduction, 
 the hadronic final state interaction in kaon photoproduction
 has usually been
neglected. Without rescattering contributions the $T$-matrix is simply
approximated by the driving term alone which is assumed to be given by 
a series of tree-level diagrams.  
Clearly, neglecting the final meson-baryon interaction in the full meson
photoproduction $T$-matrix automatically leads to violation of unitarity since flux
that can "leak out" into inelastic channels has not been properly
accounted for.  Enforcing unitarity dynamically requires
solving a system of coupled channels with all possible final states. 
In principle, this would require information
on channels, such as $K^+ \Lambda  \rightarrow K^+ \Lambda$, for which no experimental
information is available for obvious reasons. In practice,
the coupling of the channels leads to an overdetermination of the
free parameters, thus processes as the one mentioned above will be
determined along with the experimentally accessible ones.

In contrast to eta photoproduction with the dominating $S_{11}$(1535)
resonance at low energies there is no single prominent resonance in the process 
$p(\gamma, K^+)\Lambda$ at low energies. The $\pi N$ partial 
$s$- and $p$-waves do not show any cusp effect from the
opening of the $K^+ \Lambda$ threshold around $W = 1660$ MeV,
in contrast to the very pronounced cusp at the $\eta N$ threshold
visible in the $S_{11}$ partial amplitude.
The $E_{0+}$ pion photoproduction multipole on the other hand, shown in Fig. 1, 
has some structure around the $K \Lambda$ threshold
 which may also be a signal of the $S_{11}(1650)$ state. This confluence
of a resonance close to the $K^+ \Lambda$ threshold certainly appears
similar to the situation in the $\eta N$ channel.
The real part of the $M_{1-}$ multipole shows a small but clear signal
around $W = 1600$ MeV, there is no known $P_{11}$ resonance
in this energy regime.  As is obvious from Fig. 1, especially the imaginary
parts of the multipoles are not well known at higher energies; new JLab data
on pion photoproduction are expected to improve the situation.

\begin{figure}[!ht]
\centerline{\epsfysize=8.5 cm \epsfbox{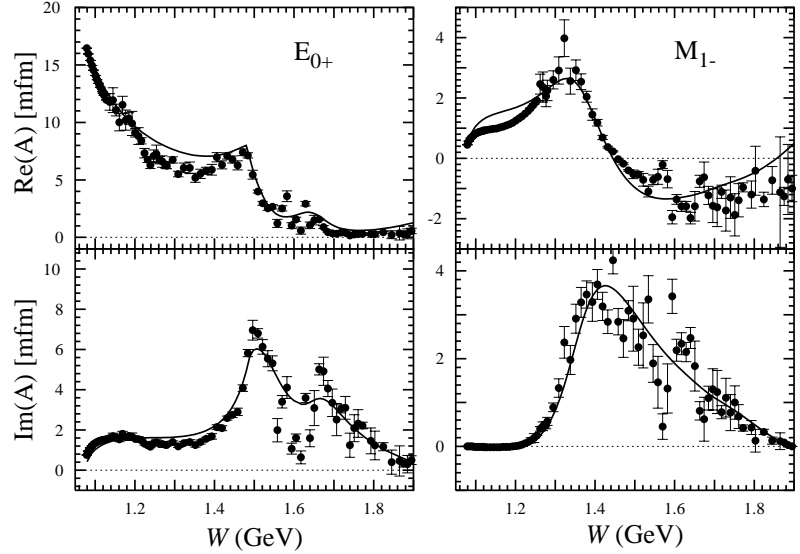}}
\vspace{-0.8cm}
\caption[thanks]{\label{fig:piphot} 
        Fits to the proton multipoles of pion photoproduction using 
        the SP97 results of the VPI group. The solid curve shows the coupled-channels result of   
                      Ref. \cite{ben:feuster98} in the K-matrix approximation.}
\end{figure}

The clearest indication of the coupling of the $K^+ \Lambda$ channel to the $\pi N$ channels
appears to come in the inelastic cross sections, shown in Fig. 2. For the $P_{11}$ channel
 the total $\pi N \rightarrow \pi \pi N$ cross section begins deviating
from the total $\pi N$ inelastic cross section at around $W = 1650$ MeV, clearly
indicating the opening of another threshold. 

\begin{figure}[!ht]
\centerline{\epsfysize=5.2 cm \epsfbox{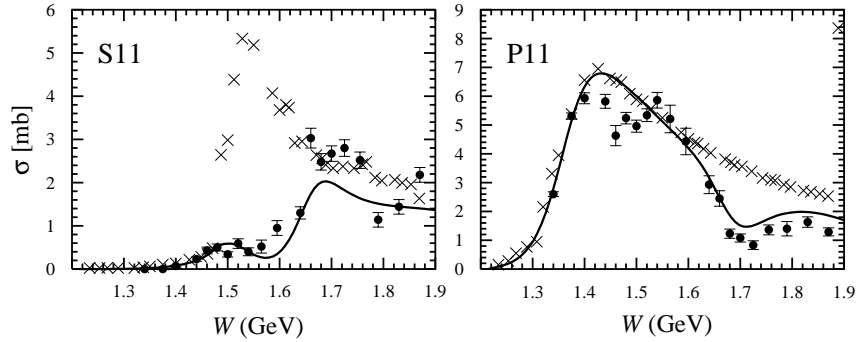}}
\caption[thanks]{\label{fig:twopi} 
        Fits to the partial $\pi N \rightarrow \pi \pi N$
        cross sections using the SM95 results of the VPI group.
        In addition, the total inelastic cross section (x)
        as determined from SM95 is shown.       
        The notation is as in Fig.1.}
\end{figure}

Proper inclusion of rescattering effects into the kaon photoproduction
process requires the accurate description of hadronic kaon production.
Among the possible rescattering reactions, only $\pi N \rightarrow
K \Lambda$ has been measured, processes such as
 $\eta N \rightarrow K \Lambda$ and $K \Lambda$ elastic scattering
have to be determined through the coupled channels approach.
Figure 3 compares the coupled-channels result to available data.
At threshold, only $s$- and $p$-waves contribute which receive 
their main contributions from the $S_{11}$(1650) and $P_{11}$(1710)
resonances which are found to decay into the $K \Lambda$ channel
with a branching ratio of about 6\% and 14\%, respectively.
The strong forward peaking at higher energies is due mostly to 
the $t$-channel $K^*$ contribution which contributes to all partial waves.
Because of the magnitude of this contribution 
the $t$-channel vector meson contributions may need to be modified
to reflect more closely a desired Regge-like behavior at high energies.

\begin{figure}[!ht]
\centerline{\epsfysize=8.3 cm \epsfbox{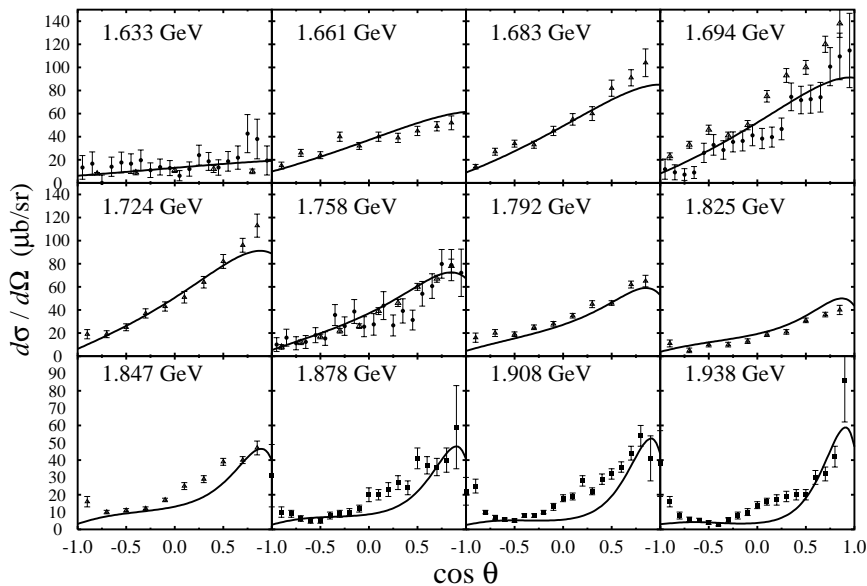}}
\caption[thanks]{\label{fig:pikalam} The coupled-channels calculation
        of Ref. \cite{ben:feuster98}
        compared to differential cross section data of the
        $\pi^- p \rightarrow K^0 \Lambda$ reaction.}
\end{figure}

Including unitarity properly also raises the question of
crossing symmetry which is straightforward to impose at the tree 
level \cite{ben:saghai,ben:williams} but more involved in a coupled-channels
framework.  This becomes
apparent when one compares the intermediate hadronic states of $p(\gamma,
K^+)\Lambda$ with those of $p(K^-,\gamma)\Lambda$. While these two processes
are related via crossing at the tree level, the photoproduction process
proceeds trough intermediate states with zero strangeness while the radiative
capture reaction requires $S=-1$. At present, no $\Lambda^*$ and $\Sigma^*$
resonances have been included in our approach in order to limit the number
of free parameters; thus crossing symmetry is violated.

\section{Hadronic Form Factors and Gauge Invariance}

\begin{figure}[!ht]
\centerline{\epsfysize=2.8 cm \epsfbox{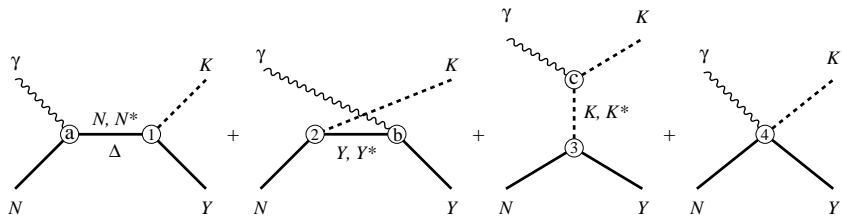}}
\caption[thanks]{\label{diagram} Feynman diagrams for the electromagnetic
  production of kaons on the nucleon. 
  Contributions from the $\Delta$ are only possible in $\Sigma$
  production. Electromagnetic vertices are denoted by (a), (b), and (c),
  hadronic vertices by (1), (2), and (3). The contact diagram (4) is
  required in both PS and PV couplings in order to restore gauge invariance
  after introducing hadronic form factors. The Born terms contain the $N$,
  $Y$, $K$ intermediate states and the contact term.}
\end{figure}

It has been well known that including hadronic form factors at hadronic
vertices in Fig.~\ref{diagram} can lead to the violation
of gauge invariance in the Born 
amplitude. Furthermore, most isobaric models show a divergence at higher
energies, which clearly demonstrates the need for a cut-off. 
 Recent calculations \cite{ben:saghai,ben:mart95} demonstrated that many 
 models which are able to
describe $(\gamma, K^+)$ experimental data tend to unrealistically overpredict
the $(\gamma, K^0)$ channel. The use of point-like particles disregards
the composite nature of nucleons and mesons, thus losing the full
complexity of a strongly interacting hadronic system

In the model of Ref. \cite{ben:mart_s96} a hadronic form factor was 
introduced by multiplying the entire photoproduction
 amplitude [see Eq.~(\ref{mfi}) below] with an overall monopole 
form factor $F(\Lambda,t) = (\Lambda^2-m_{K}^2)/(\Lambda^2-t)$,
where the cut-off mass $\Lambda$ was treated as a free parameter.   In spite of
successfully minimizing the $\chi^2$ while maintaining gauge invariance,
there is no microscopic basis for this approach. 

One method to handle the inclusion of such form factors has
been proposed by Ohta \cite{ben:ohta}. By making use of minimal substitution
Ohta has derived an additional amplitude which eliminates the form factors in
the electric terms of the Born amplitude. Recently, Haberzettl
\cite{ben:haberzettl,ben:hbmf98} has proposed another, more general method
which allows for a
multiplication of the electric terms with a form factor as well.

All amplitudes for photoproduction of pseudoscalar mesons
have the general form
\begin{eqnarray}
  \label{mfi}
  {\cal M}_{\rm f\mbox{}i}(s,t,k^2) &=& 
  {\bar u}(p_Y) \sum_{j=1}^{4} A_j(s,t,k^2)~ M_j ~ u(p_N) ~,
\end{eqnarray}
where the $M_j$ are Lorentz and gauge-invariant matrices
and the functions $A_j$ depend on kinematic variables, coupling constants and the resonance 
parameters.  The inclusion of form
factors in the hadronic vertices of the Born terms in Fig.~\ref{diagram} leads
to the modification of the four Born amplitudes $A_{j}^{\rm Born}$.
  The amplitude for each 
  resonance is separately gauge invariant, by construction.
The amplitudes for kaon photoproduction are given by
  
\begin{eqnarray}
A_{1}^{\rm Born} & = & \frac{e g_{KYN}}{s - m_{N}^{2}} \left(Q_{N} + 
\kappa_{N} \right) ~F_1(\Lambda,s) + \frac{e g_{KY N}}{u - m_{Y}^{2}} 
\left(Q_{Y}+\kappa_{Y} \right) F_2(\Lambda,u)\nonumber\\
&& + \frac{e g_{KY' N}}{u-m_{Y'}^{2}}~ \kappa_T~ \left( 1-|Q_Y|\right) 
 F_2(\Lambda,u),\\
\label{a_2}
A_{2}^{\rm Born} & = & \frac{2e g_{KY N}}{t - m_{K}^{2}} ~  
\left( \frac{Q_{N}}{s - m_{N}^{2}}+\frac{Q_{Y}}{u - m_{Y}^{2}}\right) 
{\tilde F} ~, \\
A_{3}^{\rm Born} & = & \frac{e g_{KY N}}{s - m_{N}^{2}} ~ 
\frac{\kappa_{N}}{m_{N}} ~F_1(\Lambda,s)~, \\
A_{4}^{\rm Born} & = & \frac{e g_{KY N}}{u - m_{Y}^{2}} 
~ \frac{\kappa_{Y} }{m_{Y}} ~F_2(\Lambda,u) +  \frac{2e g_{KY' N}}{u -
  m_{Y'}^{2}} ~\left( 1-|Q_Y|\right) \frac{\kappa_T 
  F_2(\Lambda,u)}{m_{Y'} + m_{Y}} ~ , \nonumber\\
\end{eqnarray} 
where $Q_{N}$ and $Q_{Y}$ denote the
charge of the nucleon and the hyperon in $+e$ unit, while $\kappa_N$,
$\kappa_Y$, and $\kappa_T$ indicate the anomalous magnetic moments of the
nucleon, hyperon, and the transition of $\Sigma^0 \Lambda$. It is
understood that $Y'=\Sigma^0 ~[\Lambda]$ for $K\Lambda ~[K\Sigma^0]$
production. The difference between the three methods can be summarized as
follows:  

\begin{eqnarray} 
\tilde{F}_{\rm overall}\quad \rm{replaces}& &
F_1,\; F_2,\; F_3, \;\;  \mbox{and} \;\; \tilde{F} 
 \mbox{ [e.g.,}\;\;\tilde{F}_{\rm overall} = F_3 (\Lambda, t)]
      ~,\\
 {\tilde F}_{\rm Ohta} &=& 1 ~, \label{ohta_ff} \\
 {\tilde F}_{\rm Haberzettl} &=& a_1F_1(\Lambda ,s) + a_2F_2(\Lambda ,u) + 
  a_3F_3(\Lambda ,t) ,\nonumber\\
  & &  ~~~ {\rm with~~} a_1+a_2+a_3= 1 ~.  
  \label{habb_ff}
\end{eqnarray}

Results of our previous calculations within 
the tree-level approximation that compare using
an overall form factor with Ohta's method 
have previously been reported in Ref.~\cite{ben:mart_s96,ben:saphir}. 
Here we compare the different methods
with the one by Haberzettl \cite{ben:hbmf98} for kaon photoproduction,
using a covariant vertex 
parameterization without any singularities on the real axis.
\begin{eqnarray}
F_i(\Lambda ,r_i) &=& \frac{\Lambda^2}{\sqrt{\Lambda^4+\left( r_i-m_i^2 
\right)^2}} ~, ~~~i=1,2,3, \label{non_sing_ff}
\end{eqnarray}
with $r_1=s, r_2=u, r_3=t$ and $m_1=m_N, m_2=m_\Lambda , m_3=m_K$.

Here we focus only on the magnitude of the
leading Born coupling constants $g_{K\Lambda N}$ and $g_{K\Sigma N}$
extracted from the photoproduction data of $K^+\Lambda$ and $K^+\Sigma^0$.
In contrast to the well-known $\pi NN$ coupling constant,
there are  serious discrepancies between values for the {\it KYN}
coupling constants extracted from electromagnetic reactions 
and those from hadronic processes which tend to be closer to accepted 
SU(3) values.

\begin{table}[!t]
    \caption{}The leading coupling constants $g_{K\Lambda N}$ 
             and $g_{K\Sigma N}$, the hadronic form factor cut-off 
             $\Lambda$, and the $\chi^2/N$ from fitting to kaon photoproduction
             data using different methods of restoring gauge 
             invariance within the tree-level approximation.
    \renewcommand{\arraystretch}{1.3}
  \begin{center}
    \begin{tabular}{llcccc}
      \hline\hline 
   form factor &  coupling & $g_{K\Lambda N}/\sqrt{4\pi}$
   & $g_{K\Sigma N}/\sqrt{4\pi}$ & $\Lambda$ (GeV) & $\chi^2/N$ \\
   method & constants & & & & \\
   \hline
   no & fixed &         $-3.80$  &   1.20        & -     & 55.76 \\
   no & free &          $-1.90$  &   $-0.37$     & -     & 3.33  \\
   overall & fixed &    $-3.80$  &   1.20        & 0.213 & 2.84  \\
   Ohta & fixed &       $-3.80$  &   1.20        & 1.422 & 14.21 \\
   Haberzettl & fixed & $-3.80$  &   1.20        & 1.128 & 4.63   \\
   SU(3) & - &   $-3.70\pm 0.70$ & 1.10$\pm$0.20 & -     & -     \\
       \hline\hline 
    \end{tabular}
    \label{tab:coupling}
  \end{center}
\end{table}

Our numerical results for the coupling constants
using the different methods are summarized in Table 
\ref{tab:coupling}, in comparison to the predictions of SU(3). 
If the leading coupling constants 
$g_{K \Lambda N} / \sqrt{4 \pi}$ and $g_{K \Sigma N} / \sqrt{4 \pi}$ 
are not allowed to vary freely and are fixed at reasonable SU(3) 
values of $-3.80$ and 1.20 (close to what is obtained from hadronic 
reactions \cite{ben:timmer}),  
the $\chi^2$ obtained in our model {\it without} hadronic form factors
comes out to be 55.8.  
On the other hand, if the two couplings are allowed
to vary freely, one obtains $g_{K \Lambda N} / \sqrt{4 \pi} = -1.90$ and
$g_{K \Sigma N} / \sqrt{4 \pi} = -0.37$ with
$\chi^2/N=3.3$. In spite of the small
$\chi^2/N$ in the latter case, this result obviously indicates that either
there is a very large amount of SU(3) symmetry breaking or that important
physics has been left out in the extraction of coupling constants from 
the $(\gamma,K)$ processes.
In Ref. \cite{ben:hbmf98}, we advocate the second position and demonstrate that 
the inclusion of structure at the hadronic vertex permits an adequate
description of kaon photoproduction with couplings close to the
SU(3) values, provided one uses the appropriate gauge procedure.

\begin{figure}[!ht]
\centerline{\epsfysize=13 cm \epsfbox{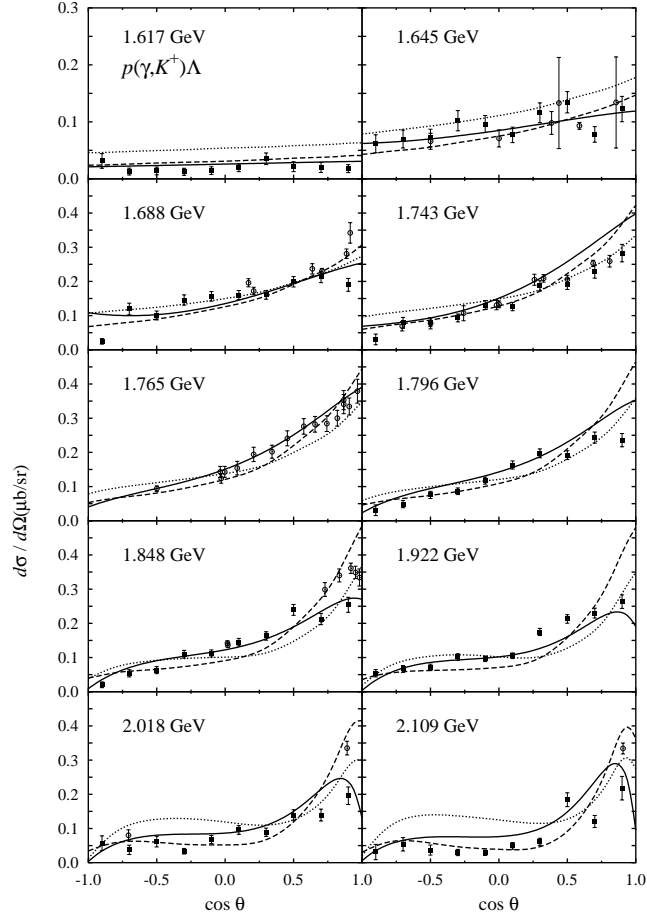}}
\caption[thanks]{\label{fig:difkpl} Differential cross section
        for the $p(\gamma,K^+)\Lambda$ channel. In the K-matrix approximation
        the dotted curves have been obtained with Ohta's prescription, 
        the dashed curves are due to
        Haberzettl's method. The solid curves show a tree-level
        calculation (Set II of Table 2) with Haberzettl's method.
        Old data are shown by open circles, new SAPHIR data
        by solid squares \cite{ben:saphir98}.}
\end{figure}

Figure 5 compares the different gauge prescriptions to the new kaon photoproduction 
data from SAPHIR \cite{ben:saphir98}.  Especially at
threshold and at higher energies, it is evident that the method of Ref. \cite{ben:hbmf98} 
is superior to the approach by Ohta.  

\section{Kaon Photoproduction in the Tree-Level Approximation}

Almost all previous analyses of kaon photoproduction 
were performed at tree level \cite{ben:saghai,ben:williams,
ben:mart95,ben:mart_s96}.
While this leads to the violation of unitarity as discussed above,
this kind of isobaric model 
provides a simple tool to parameterize meson photoproduction 
off the nucleon because it is relatively easy to calculate and to use for 
production on nuclei. 
Without rescattering contributions the $T$-matrix is simply
approximated by the driving term alone which is assumed to be given by 
a series of tree-level diagrams.
The selected Feynman diagrams 
 for the $s$-, $u$-, and $t$-channel contain some unknown coupling
parameters to be adjusted in order to reproduce experimental data. 
Final state interaction is effectively absorbed in these
coupling constants which then cannot easily be compared to couplings
from other reactions.  Guided by the coupled-channel results we have therefore
constructed a tree-level amplitude that reproduces all available $K^+ \Lambda$,
$K^+ \Sigma^0$ and $K^0 \Sigma^+$ data and thus provides an effective parameterization
of the process.  In this model, we include the three resonances that have been found
in the coupled-channels approach to decay into the $K \Lambda$ channel, the 
$S_{11}$(1650), $P_{11}$(1710), and $P_{13}(1720)$. For $K \Sigma$ production
we also allow contributions from the $S_{31}$(1900) and $P_{31}$(1910) $\Delta$ 
resonances. Furthermore, we include not only the usual $1^-$ vector meson $K^*(892)$,
but also the $1^+$ pseudovector meson $K_1(1270)$ in the $t$-channel
 since a number of studies \cite{ben:saghai,ben:williams,ben:adelseck}
  have found this resonance to give a significant contribution.

In order to describe all six isospin 
channels of $K \Lambda$ and $K \Sigma$ simultaneously 
we invoke isospin symmetry for the strong coupling constants

\begin{eqnarray}
 g_{K^{+} \Lambda p} &=& g_{K^{0} \Lambda n} ~,\\
 g_{K^{+} \Sigma^{0} p} &=& -g_{K^{0} \Sigma^{0} n}
 = g_{K^{0} \Sigma^{+} p}/ \sqrt{2} = g_{K^{+} \Sigma^{-} n}/ \sqrt{2} ~,\\
 g_{K^{+} \Sigma^{0} \Delta^{+}} &=& g_{K^{0} \Sigma^{0}\Delta^{0}} 
 ~=~ -\sqrt{2} g_{K^{0} \Sigma^{+} \Delta^{+}} ~=~ 
 \sqrt{2} g_{K^{+} \Sigma^{-} \Delta^{0}} ~.
\end{eqnarray}

The electromagnetic resonance couplings of the proton and the neutron 
are related to helicity amplitudes. For spin 1/2 resonances we have

\begin{eqnarray}
\frac{g_{N^{*0}n\gamma}}{g_{N^{*+}p\gamma}} &=&
\frac{A_{1/2}^n}{A_{1/2}^p} ~,
\end{eqnarray}
while there are two couplings for spin-3/2 resonances
\begin{eqnarray}
\frac{g^{(1)}_{N^{*0}n\gamma}}{g^{(1)}_{N^{*+}p\gamma}} &=&
\frac{\sqrt{3}A_{1/2}^n \pm A_{3/2}^n }{\sqrt{3}A_{1/2}^p \pm A_{3/2}^p } ~,\\
\frac{g^{(2)}_{N^{*0}n\gamma}}{g^{(2)}_{N^{*+}p\gamma}} &=&
\frac{\sqrt{3}A_{1/2}^n - ({m_N}/{m_{N^*}}) A_{3/2}^n }{\sqrt{3}A_{1/2}^p 
 - ({m_N}/{m_{N^*}}) A_{3/2}^p } ~
\end{eqnarray}

The helicity amplitudes quoted in the Particle Data Tables \cite{ben:pdg98}
have large error bars, especially for the neutron values.
We use the following values for the ratios: 
$g_{S_{11}(1650) n \gamma}$/$g_{S_{11}(1650) p \gamma} = -0.28$, 
$g_{P_{11}(1710) n \gamma}/g_{P_{11}(1710) p \gamma} = -0.22$,
and 
$g^{(1)}_{P_{13}(1720) n \gamma}/g^{(1)}_{P_{13}(1720) p \gamma} = -2.24$,
$g^{(2)}_{P_{13}(1720) n \gamma}/g^{(2)}_{P_{13}(1720) p \gamma} = 0.42$
for the various resonances.

For neutral kaon photoproduction the transition moments $g_{K^{*+}K^+\gamma}$
in $K^+$ photoproduction must be replaced by the neutral transition moment
$g_{K^{*0}K^0\gamma}$. The transition moment is
related to the decay width which are well known for the $K^*(892)$, i.e.

\begin{eqnarray}
\Gamma_{K^{*+}\rightarrow K^+\gamma} &=& 50\pm 5 ~{\mathrm keV} ~,\\
\Gamma_{K^{*0}\rightarrow K^0\gamma} &=& 117\pm 10 ~{\mathrm keV} ~.
\end{eqnarray}

Thus, we obtain $g_{K^{*0}K^0\gamma} = -1.53~ g_{K^{*+}K^+\gamma}$,
where we have used a quark model prediction to constrain the relative sign.

The decay widths of $K_1(1270)$, on the other hand, are not well known.
We therefore take
the ratio of the charged to neutral moment of the $K_1(1270)$ 
as a free parameter that is fixed by data in the $p(\gamma,K^0)\Sigma^+$ channel.

In order to approximately account for unitarity corrections
at tree-level we include energy-dependent widths in the resonance propagators 

\begin{eqnarray}
\Gamma ({\mbox{\boldmath ${q}$}}) &=& \Gamma_{N^*}~
\frac{\sqrt{s}}{m_{N^*}}~
  \sum_{i} x_i \left( 
\frac{|{\mbox{\boldmath ${q}$}}_i|}{|{\mbox{\boldmath ${q}$}}_i^{N^*}|}
  \right)^{2l+1} \frac{D_l(|
  {\mbox{\boldmath ${q}$}}_i|)}{D_l(|{\mbox{\boldmath ${q}$}}_i^{N^*}|)} ~,
\label{eq:decay_width}
\end{eqnarray}
where the sum runs over the possible decay channels into a meson and 
a baryon with mass $m_i$ and $m_b$, respectively, and relative orbital angular
momentum $l$. In Eq.~(\ref{eq:decay_width}), 
$\Gamma_{N^*}$ represents the total decay width and 
$x_i$ denotes the relative branching ratio of the resonance into the $i$th channel.
The final state momenta are given by  
$|{\mbox{\boldmath {q}}}_i^{N^*}|$ =
 $ \left[(m_{N^*}^2-m_b^2+m_i^2)^2/4m_{N^*}^2 -m_i^2\right]^{1/2} ~$,
and
$|{\mbox{\boldmath {q}}}_i|$ = $\left[(s-m_b^2+m_i^2)^2/4s
  -m_i^2\right]^{1/2} ~$, 
while the fission barrier factor $D_l({\mbox{\boldmath ${q}$}})$ is chosen as
$ D_l({\mbox{\boldmath ${q}$}}) = {\mathrm exp} \left( -
{\mbox{\boldmath ${q}$}}^2/3\alpha^2 \right)~$
with $\alpha = 410 $ MeV.


\begin{table}
\caption{}
Coupling constants fitted in our model. Set I results from using our 
previous model which fits old photo- and electroproduction data
\protect\cite{ben:mart_s96}, set II shows the result of this
study. Note that we have used the following parameterization of the 
form factor of Eq. (14):
${\tilde F} = \sin^2\Theta_{\rm hd} \cos^2\Phi_{\rm hd} F(\Lambda,s) +
\sin^2\Theta_{\rm hd}\sin^2\Phi_{\rm hd}F(\Lambda,u) + 
\cos^2\Theta_{\rm hd} F(\Lambda,t) $,
where the combination of sinusoidal functions ensures the correct
normalization of the form factor. Both $\Theta_{\rm hd}$ and 
$\Phi_{\rm hd}$ are obtained from the fit. $\Lambda_1$ refers 
to the cut-off for the background terms while $\Lambda_2$ denotes 
the resonance cut-off.

\begin{flushleft}        
\renewcommand{\arraystretch}{1.4}
\label{tab:cc}
\begin{center}
\begin{tabular}{lrr}
\hline\hline
Coupling constants & Set I & Set II\\
[0.5ex]
\hline
$g_{K\Lambda N}/\sqrt{4\pi}$ & $-3.094\pm 0.077$ & $-3.800$  \\
$g_{K\Sigma N}/\sqrt{4\pi}$ & $1.227\pm 0.055$ & $1.200$   \\
$\Theta_{\rm hd}~ (^\circ)$ & -~~ & $108\pm 4$\\
$\Phi_{\rm hd}~ (^\circ)$ & -~~ & $90\pm 6$\\
$\Lambda_{1}$ (GeV) & $0.853\pm 0.018$ & $0.798\pm 0.006$\\
$\Lambda_{2}$ (GeV) & -~~ & $1.883\pm 0.110$\\
[1ex]
\hline
$K\Lambda$ coupling \\
[0.5ex]
\hline
$g_{K^*K\gamma}~g^V_{K^*\Lambda N}/{4\pi}$&$-0.188\pm 0.006$&
$-0.506\pm 0.013$ \\
$g_{K^*K\gamma}~g^T_{K^*\Lambda N}/{4\pi}$ &$-0.122 \pm 0.018$& 
$ 0.672\pm 0.065$ \\
$g_{K_1K\gamma}~g^V_{K_1\Lambda N}/{4\pi}$ &-& $ 0.063\pm 0.073$ \\
$g_{K_1K\gamma}~g^T_{K_1\Lambda N}/{4\pi}$ &-& $ 0.372\pm 0.209$ \\
$g_{N^*(1650)N\gamma}~g_{K\Lambda N^*(1650)}/\sqrt{4\pi}$ & ~~~
$-0.063 \pm 0.005$&~~~ $-0.130\pm 0.001$ \\
$g_{N^*(1710)N\gamma}~g_{K\Lambda N^*(1710)}/\sqrt{4\pi}$ &
$-0.065 \pm 0.019$& $-0.094\pm 0.011$ \\
$g^{(1)}_{N^*(1720)N\gamma}~g_{K\Lambda N^*(1720)}/\sqrt{4\pi}$ &
- & $0.060\pm 0.003$ \\
$g^{(2)}_{N^*(1720)N\gamma}~g_{K\Lambda N^*(1720)}/\sqrt{4\pi}$ &
-& $0.943\pm 0.021$ \\
[1ex]
\hline
$K\Sigma$ coupling \\
\hline
$g_{K^*K\gamma}~g^V_{K^*\Sigma N}/{4\pi}$ &$-0.079 \pm 0.005$& 
$-0.306\pm 0.013$\\
$g_{K^*K\gamma}~g^T_{K^*\Sigma N}/{4\pi}$ &$-0.079 \pm 0.020$& 
$-0.603\pm 0.017$\\
$g_{K_1K\gamma}~g^V_{K_1\Sigma N}/{4\pi}$ &- & $-0.397\pm 0.038$\\
$g_{K_1K\gamma}g^T_{K_1\Sigma N}/{4\pi}$ &- & $-1.710\pm 0.216$\\
$g_{N^*(1650)N\gamma}~g_{K\Sigma N^*(1650)}/\sqrt{4\pi}$ &
$-0.007 \pm 0.015$& $-0.041\pm 0.003$\\
$g_{N^*(1710)N\gamma}~g_{K\Sigma N^*(1710)}/\sqrt{4\pi}$ &
$2.100 \pm 0.102$& $0.084\pm 0.018$\\
$g_{\Delta(1900)N\gamma}~g_{K\Sigma\Delta(1900)}/\sqrt{4\pi}$ &
$0.234 \pm 0.015$&$0.104\pm 0.002$\\
$g_{\Delta(1910)N\gamma}~g_{K\Sigma\Delta(1910)}/\sqrt{4\pi}$ &
$-0.991 \pm 0.091$&$0.363\pm 0.017$\\
$g_{K_1^0K^0\gamma}~/~g_{K_1^+K^+\gamma}$ &- & $0.261\pm 0.210$ \\
[1ex]
\hline
$\chi^2/N$ &5.99 & 3.45  \\
[0.5ex]
\hline\hline
\end{tabular}
\end{center}
\end{flushleft}
\end{table}

We have performed a combined fit to all differential
cross section and  recoil polarization data of 
$p(\gamma,K^+)\Lambda$ and $p(\gamma,K^+)\Sigma^0$.
The $p(\gamma,K^0)\Sigma^+$ channel is included later, since data for this 
channel have large error bars, and therefore do not strongly influence
the fit. The results of our fits are summarized in Table 2.
We compare our present study to an older model \cite{ben:mart_s96}
which employed an overall hadronic form factor and did not contain
the $P_{13}$(1720) and the $K_1$(1270) states. The significant improvement
in $\chi^2$ comes mostly from including the $P_{13}$(1720) in
the $K \Lambda$ channel; its decay into 
the $K \Sigma$ channel is negligible.  A further reduction in $\chi^2$
results from allowing the non-resonant background terms to have a different
form factor cut-off than the $s$-channel resonances.  For the former, the fit produced
a soft value of about 800 MeV, leading to a strong suppression of the 
background terms while the resonance cut-off is determined to be 1.89 GeV.
This combination leads to a reaction mechanism
 which is resonance dominated in all isospin channels.
Table 2 reveals that the coupling
ratio $K_1^0K^0\gamma / K_1^+K^+\gamma$ is obtained with 
large uncertainty. This comes as no surprise since the data
in the $p(\gamma,K^0)\Sigma^+$ channel have large error-bars;
we predict the ratio of the decay widths to be
\begin{eqnarray}
\frac{\Gamma_{K_1^0\rightarrow K^0\gamma}}{\Gamma_{K_1^+\rightarrow
K^+\gamma}} &=& 0.068 \pm 0.110 ~.
\end{eqnarray}

The differential cross section of 
$p(\gamma,K^+)\Lambda$ was already shown in Fig. 5, where we compare
the tree-level fit (Set II in Table 2) with coupled-channel results.
At threshold, the process is dominated by $s$-wave, coming from the
$S_{11}$(1650) state. At higher energies we find strong forward
peaking similar to the $p(\pi^-,K^0)\Lambda$ case
that can again be attributed to the $K^*$ contribution.

\begin{figure}[!ht]
\centerline{\epsfysize=9 cm \epsfbox{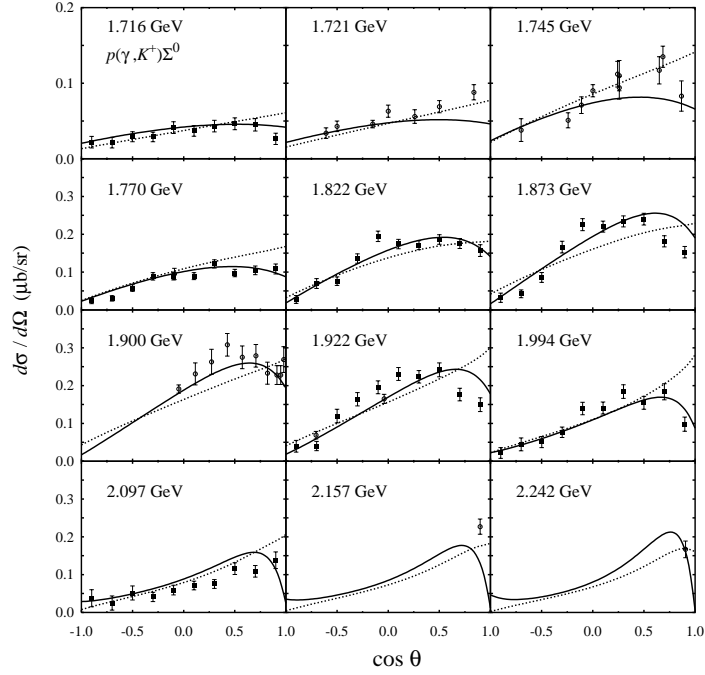}}
\caption[thanks]{\label{fig:difkps0} Differential cross section for $p(\gamma ,K^+)\Sigma^0$ channel
    calculated at tree level. The solid curve shows Set II of Table 2
    while the dashed line shows the older model, Set I of Table 2.}
\end{figure}

The comparison of the two models with the $p(\gamma,
K^+)\Sigma^0$ data is shown in Fig.~\ref{fig:difkps0} 
from threshold up to 2.2 GeV. In contrast to $K^+ \Lambda$
photoproduction, this channel contains significant $p$- and $d$-wave contributions
already at threshold. This points to the $P_{11}$(1710) state as an important
resonance in low-energy $K \Sigma$ production; here the $S_{11}$(1650) 
lies below threshold. This finding is consistent with a recent 
study \cite{ben:sibirtsev98}
of $K \Sigma$ production in NN scattering, $N N \rightarrow N K \Sigma$,
where the $P_{11}$(1710) state was identified as a major contribution.

\begin{figure}[!ht]
\centerline{\epsfysize=5 cm \epsfbox{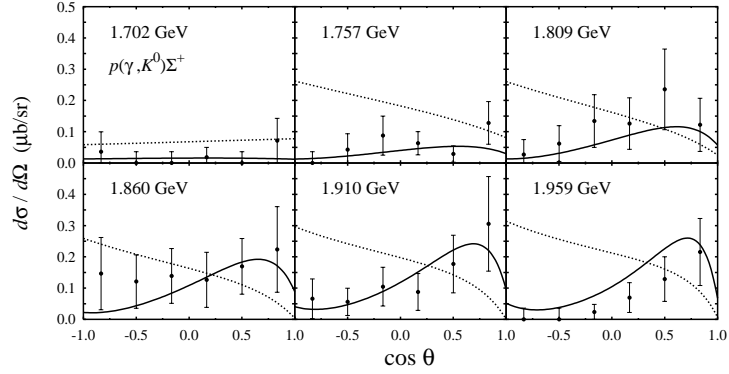}}
\caption[thanks]{\label{fig:difk0sp} Differential cross section for $p(\gamma ,K^0)\Sigma^+$ channel.
             Data are from Ref.~\protect\cite{ben:saphir}.
             Notation is as in Fig. \ref{fig:difkps0}.}
\end{figure}

Figure \ref{fig:difk0sp} compares the two models for the
$p(\gamma,K^0)\Sigma^+$ channel. The new SAPHIR data are
clearly able to discriminate between the models. The model corresponding to 
Set I not only overpredicts the data at threshold, but also yields a backward
peaking behavior, while the data tend to favour forward peaking.
This dramatically different behavior is due mostly to the different
gauge prescriptions used since this influences the relative contribution
of the background terms.
As mentioned above, Set I used an overall hadronic form factor that 
multiplied the entire amplitude, while Set II employs 
the mechanism by Haberzettl, which is clearly preferred by the data.

\begin{figure}[!ht]
\centerline{\epsfysize=11 cm \epsfbox{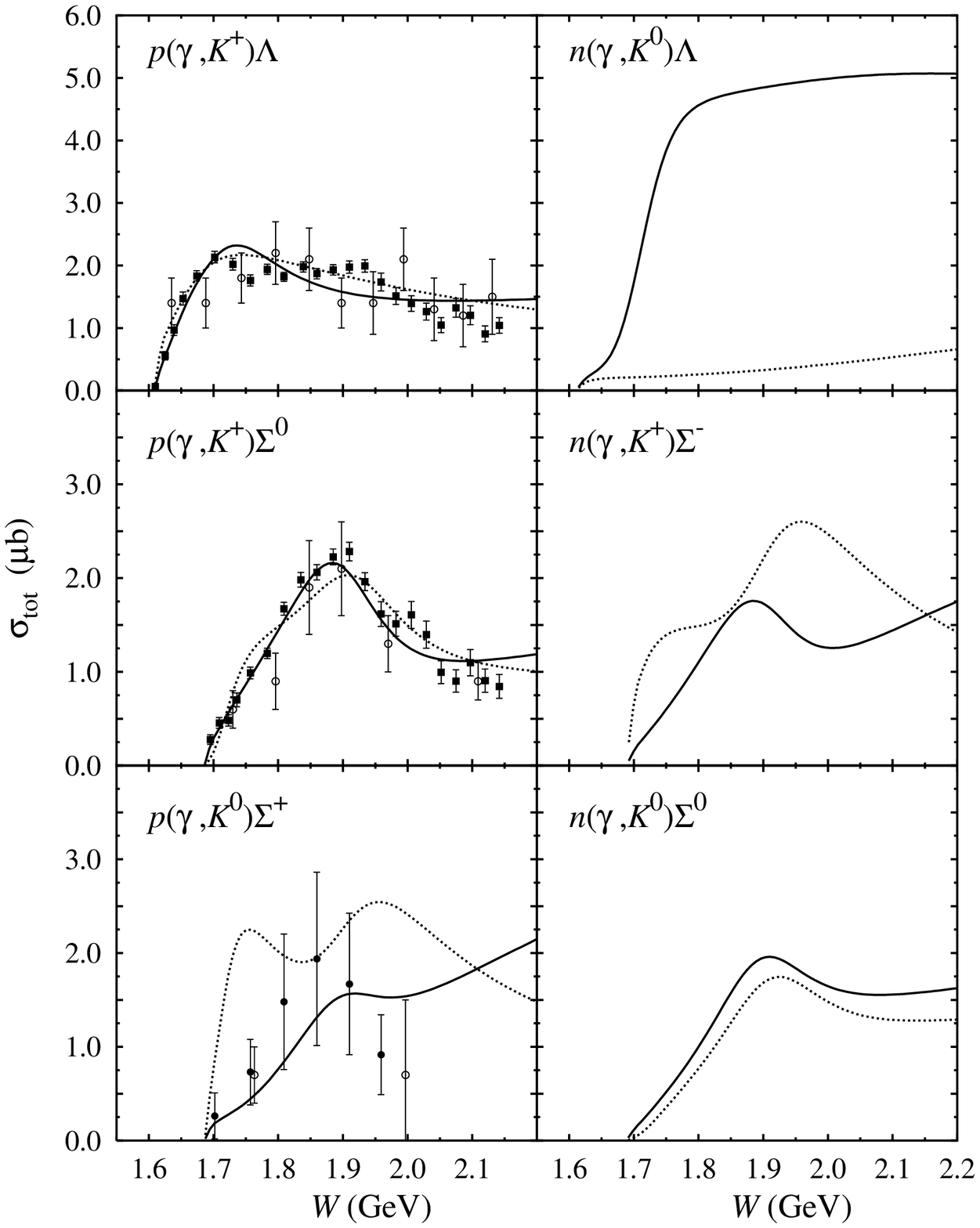}}
\caption[thanks]{\label{fig:total} Total cross sections
        for the six isospin channels
        of kaon photoproduction on the nucleon. The data for $K^+$ production 
        come from Ref. \cite{ben:saphir98} and the data for $K^0$ production are
        from Ref. \cite{ben:saphir}.
        The notation of the curves is as in Fig. 6.}
\end{figure}

Figure \ref{fig:total} compares total cross section data for the three different
$K^+$ photoproduction reactions on the proton. For $p(\gamma,K^+)\Lambda$
one can clearly see the cusp effect around $W = 1710$ MeV, indicating
the opening of the $K \Sigma$ channel. The steep rise of the $K^+ \Lambda$
data at threshold is again indicative of a strong $s$-wave. 
The $K^+ \Sigma^0$ data rise more slowly at threshold,
suggesting $p$- and $d$-wave, rather than $s$-wave, dominance.  
Furthermore, there is a clear evidence for
a resonance structure
around $W = 1900$ MeV.
There is indeed a cluster of six or seven $\Delta$ resonances with spin
quantum numbers 1/2, 3/2, 5/2 and 7/2; it is at this energy
that the total $K \Sigma$ cross section reaches its maximum.
Coupled-channels calculations for the
 $K \Sigma$ reaction are in progress and will be reported elsewhere.

\begin{figure}[!t]
\centerline{\epsfysize=8 cm \epsfbox{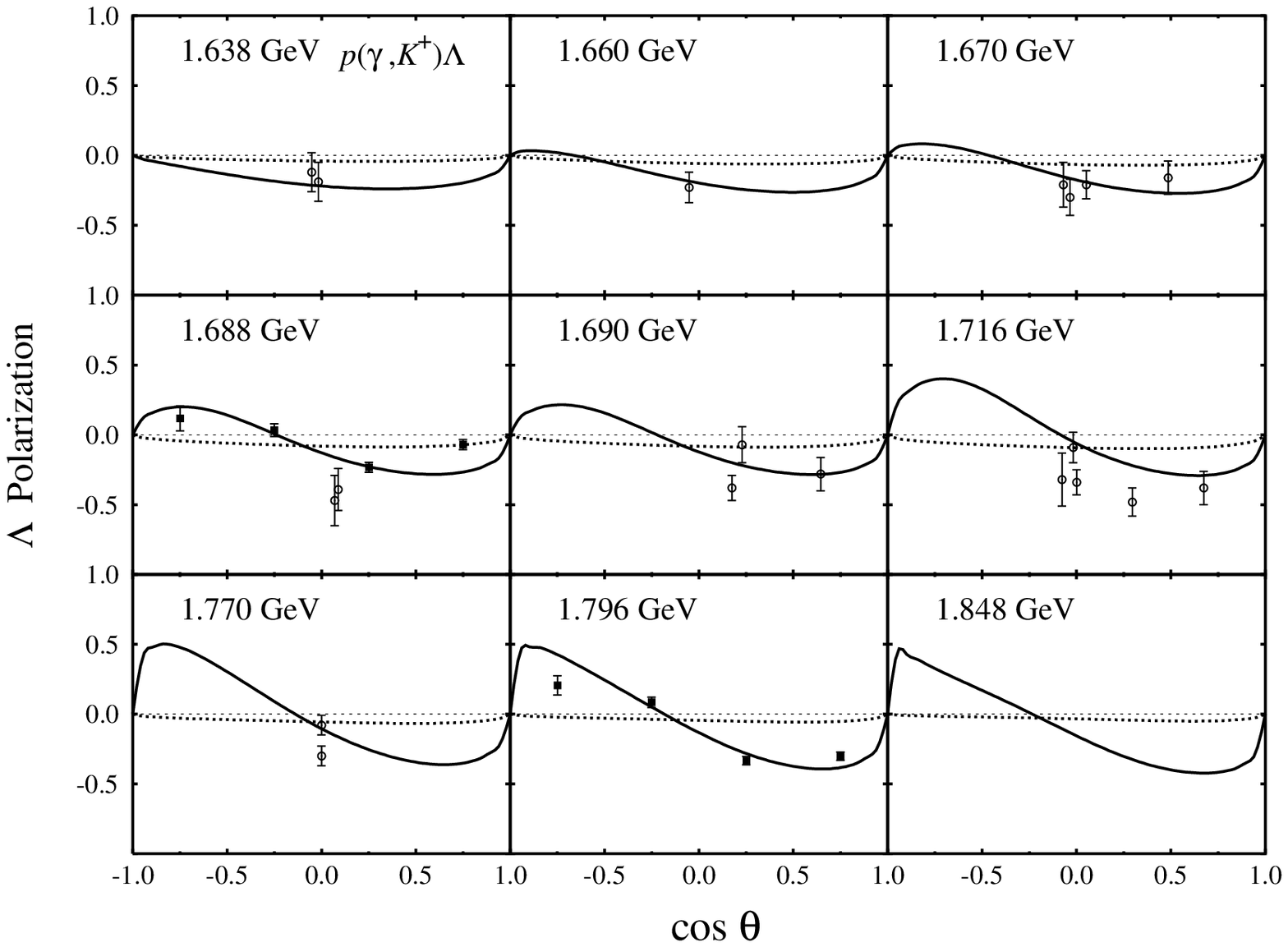}}
\caption{\label{fig:pollam}$\Lambda$ recoil polarization for $p(\gamma ,K^+)\Lambda$.
             Notation is as in Fig. \ref{fig:difkps0}.}
\end{figure}

\begin{figure}[!t]
\centerline{\epsfysize=8 cm \epsfbox{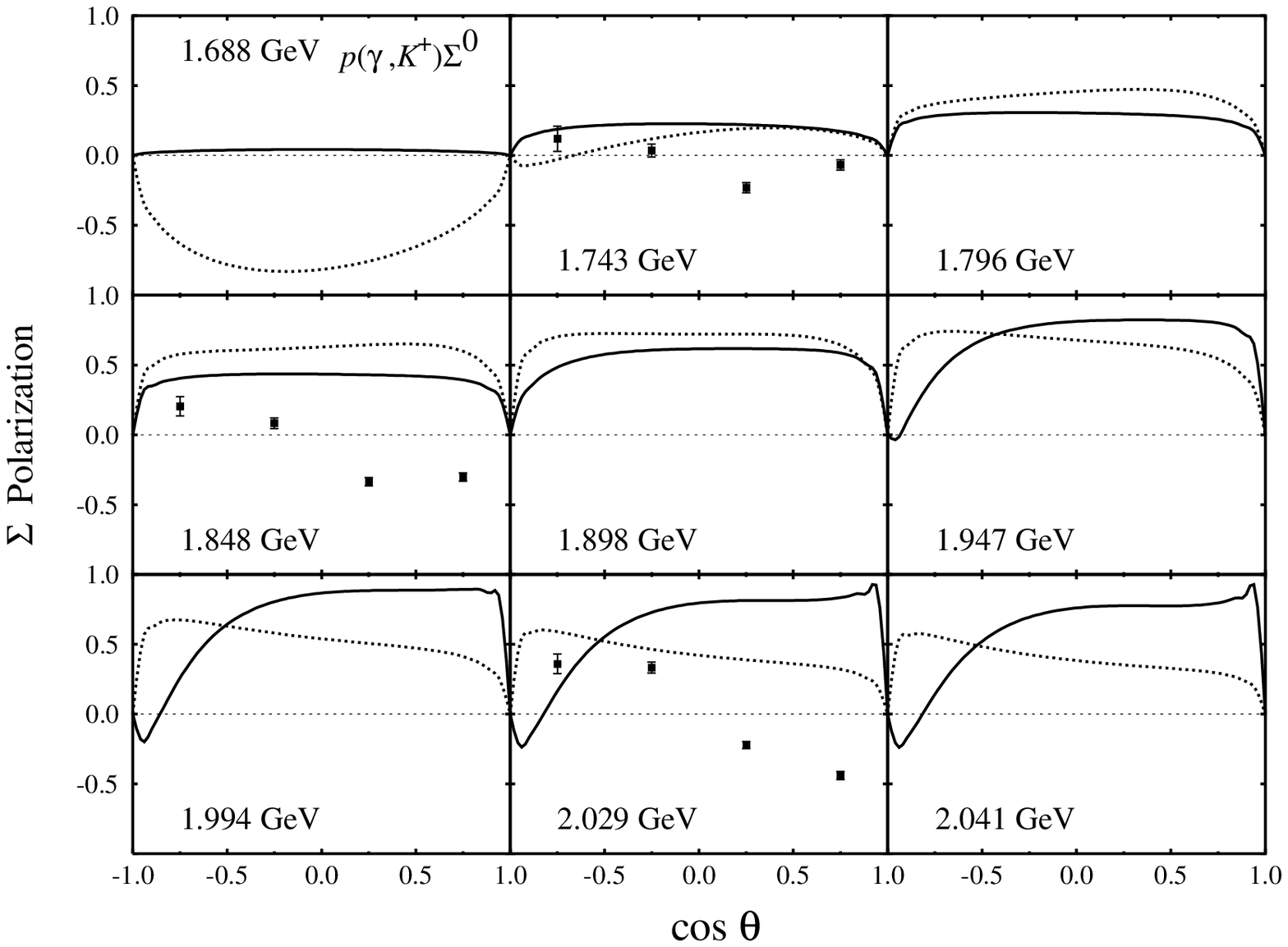}}
\caption[thanks]{\label{fig:polsig}$\Sigma^0$ recoil polarization of $p(\gamma ,K^+) \Sigma^0$.
             Notation is as in Fig. \ref{fig:difkps0}.}
\end{figure}

The recoil polarization for $K^+ \Lambda$ and $K^+ \Sigma^0$ production
is shown in Fig.~\ref{fig:pollam} and Fig.~\ref{fig:polsig}, respectively.
We find good agreement with the data using Set II of Table 2, while the older
model (Set I) gives almost zero polarization throughout this energy range.
The main reason for this dramatic difference is the more prominent role
that the resonances play in the present model, defined by Set II.
However, this model utterly fails to reproduce the polarization data
for $K^+ \Sigma^0$ production. Since the recoil polarization observable
is sensitive especially to the imaginary parts of the amplitudes this discrepancy
suggests that we do not have the correct resonance input for the $K \Sigma$ channel.

\section{Kaon Photoproduction: a Signal for "Missing" Resonances?}

A recent quark model study of Capstick and Roberts \cite{ben:capstick98}
finds that a number of missing and undiscovered nucleon resonances
have substantial decay amplitudes into the $K \Lambda$ and $K \Sigma$
final states. They conclude that electromagnetic and hadronic kaon production
in the energy region of 1.8 - 2.3 GeV
can provide a useful tool to identify new states and help to extract resonance 
parameters of weakly established states. Most of these resonances appear
to be negative-parity states with wave functions predominantly in the $N = 3$
band. 

The new SAPHIR total cross section data in the $K^+ \Lambda$ channel (shown in Fig. 8)
reveal an interesting structure around $W=1900$ MeV.  Our model fits currently
do not reproduce this bump since there is no well-established (3- and 4-star)
 I=1/2 state at this energy.                                                        
However, Ref. \cite{ben:capstick98} predicts a missing $D_{13}$ at 1960 MeV that has a
large branching ratio into the $K \Lambda$ channel. In order to 
study this structure more closely, we have included 
a $D_{13}$ resonance into our Set II model but have allowed the mass and the width of
the state to vary as free parameters. We achieve a significant 
reduction in $\chi^2 / N$ for a mass of 1902 MeV and a total width of 315 MeV.
While this clearly cannot yet be regarded as proof for the existence
of this state it nevertheless demonstrates the potential of kaon production
as a tool to discover missing nucleon resonances.

This work was supported by DOE grant DE-FG02-95ER-40907 (CB,AW, and HH),
a University Research for Graduate Education (URGE) grant (TM), 
BMBF, DFG and GSI Darmstadt (GP, TF, and UM).

\end{document}